\documentclass[showpacs,prd]{revtex4}
\usepackage{epsfig}
\usepackage{graphicx}
\usepackage{dcolumn}
\usepackage{amsmath}
\usepackage{latexsym}

\begin{document} 


\title{Free particle on noncommutative plane -- a coherent state
path integral approach}  
\date{\today}
\author{Sunandan Gangopadhyay$^{a}$\footnote{e-mail: sunandan.gangopadhyay@gmail.com, sunandan@sun.ac.za}, 
Frederik G Scholtz $^{a,b}$\footnote{e-mail:
fgs@sun.ac.za}}
\affiliation{$^a$National Institute for Theoretical Physics (NITheP), 
Stellenbosch 7600, South Africa\\
$^b$Institute of Theoretical Physics, 
University of Stellenbosch, Stellenbosch 7600, South Africa}

\begin{abstract}
\noindent We formulate the coherent state path integral on a two
dimensional noncommutative plane using the fact 
that noncommuative quantum
mechanics can be viewed as a quantum system on the 
Hilbert space of Hilbert-Schmidt operators 
acting on noncommutative configuration space.  The propagation kernel for the free 
particle shows ultra-violet cutoff which agrees with the earlier
investigations made in the literature but the approach differs
substantially from the earlier studies.
\end{abstract}

\maketitle

\noindent {\it{Introduction }}:\\

\noindent The idea of noncommutative spacetime was first formally
introduced by Snyder in \cite{snyder} as an attempt to regulate the
divergences of quantum field theories. These ideas were taken seriously when considerable evidence came from string theory
\cite{wit} to the issues of quantum gravity that suggests
that attempts to unify gravity and quantum mechanics will ultimately
lead to a noncommutative geometry of spacetime. Thereafter, 
despite a number of investigations into the possible physical consequences of noncommutativity in quantum mechanics and quantum mechanical many-body systems \cite{duval}-\cite{khan}, quantum electrodynamics 
\cite{chai}-\cite{lia}, the standard model \cite{ohl} and cosmology  \cite{gar}, \cite{alex}, our understanding of the physical implications of noncommutativity is still far from being
complete. The difficulty in having a through understanding of the physical implications of noncommutativity, is the lack of a systematic formulation and interpretational framework of noncommutative quantum mechanics. The difficulty persists even in the 
path integral formulation of noncommutative quantum mechanics
and there seems to be a lot of disagreement in the results
obtained in the literature.

\noindent The path integral formulation, in general, attempt 
to evaluate noncommutative analogues of the Feynman kernel :
\begin{eqnarray}
K(q,t;q_0,t_0)=\langle q|\hat{U}(t,t_0)|q_0\rangle
\label{intro1}
\end{eqnarray}
where, $\hat{U}$ is the unitary time evolution operator and
$|q\rangle\equiv|q_1,...,q_d\rangle$ are the position eigenkets
in $d$ dimensions. Noncommutative geometry implies the absence of
common position eigenkets. This problem was circumvented in
\cite{spallucci} by taking coherent states to define the 
propagation kernel. The coherent states being the eigenstates
of complex combinations of the position operators act as a
meaningful replacement for the position eigenstates admissible
only in the commutative theory. The observation 
made in this paper is that the free particle propagator turns
out to exhibit an ultra-violet cutoff induced by the noncommutative
parameter $\theta$. In another paper \cite{tan}, the same trick
of using coherent states to define the propagation kernel
was employed, however, the final expressions for the propagator
and the resulting physics were quite different from \cite{spallucci}. 

\noindent In this note, we develop an unambiguous formulation of the
Feynman path integral representation for the free particle propagator
using the ideas of a series of very recent papers 
by one of the authors \cite{scholtz, laure} where a full fledged formulation and interpretation
of noncommutative quantum mechanics have been carried out explicitly.
The result indeed shows the presence of a damping exponential term 
induced by the parameter of noncommutativity $\theta$
in the free particle propagator. However, in contrast
to \cite{spallucci}, the star product enters automatically in the computation once the completeness relation for the 
states living in the quantum Hilbert space are introduced. 
This is the new point in our paper which says that a systematic
formulation of noncommutative quantum mechanics automatically
brings in the star product into the game and furthermore, it is
even possible to make exact analytical computations using the star 
product.

\noindent In the next section, we briefly review the formalism
developed in \cite{scholtz, laure} to deal with noncommutative quantum systems and then derive the completeness relations for
the momentum and position eigenstates living 
in the quantum Hilbert space. Using this, we move on to construct
the path integral representation for the propagator of the 
free particle in the two dimensional noncommutative space.
In the rest of the paper, we shall work with natural units
$\hbar=c=1$.\\


\noindent {\it{Formalism and the free particle propagator}} :\\

\noindent Before constructing the path integral representation
of the free particle propagation kernel on noncommutative space, we present a brief review of the formalism of noncommutative quantum mechanics developed recently in \cite{scholtz, laure}. 
This formalism have been developed in complete analogy with commutative quantum mechanics.
 
\noindent We start by giving precise meaning to the concepts
of the classical configuration space and the Hilbert space
of a noncommutative quantum system. The first step is to define 
classical configuration space. In two dimensions, 
the coordinates of noncommutative configuration space satisfy the commutation relation 
\begin{equation}
[\hat{x}, \hat{y}] = i\theta
\label{1}
\end{equation} 
where without loss of generality it is assumed that $\theta$ is a real positive parameter. Using this, it is convenient to define the creation and annihilation operators
\begin{eqnarray}\nonumber
b = \frac{1}{\sqrt{2\theta}} (\hat{x}+i\hat{y})\quad,\quad
b^\dagger =\frac{1}{\sqrt{2\theta}} (\hat{x}-i\hat{y})
\label{2}
\end{eqnarray}
that satisfy the Fock algebra $[ b ,b^\dagger ] = 1$. 
The noncommutative configuration space is then 
isomorphic to the boson Fock space
\begin{eqnarray}
\mathcal{H}_c = \textrm{span}\{ |n\rangle= 
\frac{1}{\sqrt{n!}}(b^\dagger)^n |0\rangle\}_{n=0}^{n=\infty}
\label{3}
\end{eqnarray}
where the span is take over the field of complex numbers.

\noindent The next step is to introduce the Hilbert space
of the noncommutative quantum system. We consider the set of Hilbert-Schmidt operators acting on noncommutative configuration space
\begin{equation}
\mathcal{H}_q = \left\{ \psi(\hat{x},\hat{y}): 
\psi(\hat{x},\hat{y})\in \mathcal{B}
\left(\mathcal{H}_c\right),\;
{\rm tr_c}(\psi^\dagger(\hat{x},\hat{y})
\psi(\hat{x},\hat{y})) < \infty \right\}.
\label{4}
\end{equation}
Here ${\rm tr_c}$ denotes the trace over noncommutative 
configuration space and $\mathcal{B}\left(\mathcal{H}_c\right)$ 
the set of bounded operators on $\mathcal{H}_c$. 
This space has a natural inner product and norm 
\begin{equation}
\left(\phi(\hat{x}, \hat{y}), \psi(\hat{x},\hat{y})\right) = 
{\rm tr_c}(\phi(\hat{x}, \hat{x})^\dagger\psi(\hat{x}, \hat{y}))
\label{inner}
\end{equation}
and forms a Hilbert space \cite{hol}. This space is the analog of the space of square integrable wave functions of commutative quantum mechanics and to distinguish it from the noncommutative configuration space $\mathcal{H}_c$, which is also a Hilbert space, we shall refer to it as quantum Hilbert space and use the subscripts $c$ and $q$ to make this distinction. Furthermore, we denote states in the noncommutative configuration space by $|\cdot\rangle$ and states in the quantum Hilbert space by $\psi(\hat{x},\hat{y})\equiv |\psi)$ and the elements of its dual (linear functionals) by $(\psi|$, which maps elements of $\mathcal{H}_q$ onto complex numbers by $\left(\phi|\psi\right)=\left(\phi,\psi\right)={\rm tr_c}\left(\phi(\hat{x}, \hat{y})^\dagger
\psi(\hat{x}, \hat{y})\right)$. We also need to be careful when denoting hermitian conjugation to distinguish between these two spaces. We use the notation $\dagger$ to denote hermitian conjugation on noncommutative configuration space and the notation $\ddagger$ for hermitian conjugation on quantum Hilbert space.\\  

\noindent We now replace the abstract Heisenberg algebra  
by the noncommutative Heisenberg algebra. 
In two dimensions this reads 
\begin{eqnarray}
\label{hei}
\left[\hat{x},\hat{p}_x\right]&=& i\quad,
\quad\left[\hat{y},\hat{p}_y\right]= i\nonumber\\
\left[\hat{x}, \hat{y}\right]&=&i\theta\\
\left[\hat{p}_x,\hat{p}_y\right]&=& 0\nonumber.
\end{eqnarray}
A unitary representation of this algebra in terms of operators 
$\hat{X}$, $\hat{Y}$, $\hat{P}_x$ and $\hat{P}_y$ 
acting on the states of the quantum Hilbert space 
(\ref{4}) with inner product (\ref{inner}), 
which is the analog of the Schr\"{o}dinger 
representation of the Heisenberg algebra, is easily found to be 
\begin{eqnarray}
\label{schnc}
\hat{X}\psi(\hat{x},\hat{y}) &=& \hat{x}\psi(\hat{x},\hat{y})\nonumber\\
\hat{Y}\psi(\hat{x},\hat{y}) &=& \hat{y}\psi(\hat{x},\hat{y})\nonumber\\
\hat{P}_x\psi(\hat{x},\hat{y}) &=& \frac{1}{\theta}[\hat{y},\psi(\hat{x},\hat{y})]= -i\frac{\partial\psi(\hat{x},\hat{y})}{\partial\hat{x}}\nonumber\\
\hat{P}_y\psi(\hat{x},\hat{y}) &=& -\frac{1}{\theta}[\hat{x},\psi(\hat{x},\hat{y})]= -i\frac{\partial\psi(\hat{x},\hat{y})}{\partial\hat{y}}~.
\label{action}
\end{eqnarray}
Note that the position acts by left multiplication and the momentum adjointly.  We use capital letters to distinguish operators acting on 
quantum Hilbert space from those acting on 
noncommutative configuration space. 
It is also useful to introduce the following quantum operators 
\begin{eqnarray}
\label{qop}
B&=&\frac{1}{\sqrt{2\theta}}\left(\hat{X}+i\hat{Y}\right)\nonumber\\
B^\ddagger&=&\frac{1}{\sqrt{2\theta}}\left(\hat{X}-i\hat{Y}\right)\nonumber\\
\hat{P}&=&\hat{P}_x + i\hat{P}_y\nonumber\\
\hat{P}^\ddagger &=& \hat{P}_x -i\hat{P}_y~.
\label{operators}
\end{eqnarray}
We note that $\hat{P}^2=\hat{P}^2_x+\hat{P}^2_y 
= P^\ddagger P = PP^\ddagger$. These operators act in the following way
\begin{eqnarray}
\label{action1}
B\psi(\hat{x},\hat{y}) &=& b\psi(\hat{x},\hat{y})\nonumber\\
B^\ddagger\psi(\hat{x},\hat{y}) &=& b^\dagger\psi(\hat{x},\hat{y})\nonumber\\
P\psi(\hat{x},\hat{y})&=& -i \sqrt{\frac{2}{\theta}}[b,\psi(\hat{x},\hat{y})]
\nonumber\\
P^\ddagger\psi(\hat{x},\hat{y}) &=& i
\sqrt{\frac{2}{\theta}}[ b^{\dagger},\psi(\hat{x},\hat{y})].
\end{eqnarray}
The operator $\psi(\hat{x},\hat{y})$ is just a vector in the quantum
Hilbert space and can also be denoted as 
$|\psi)=\psi(\hat{x},\hat{y})$.
With the above notions in place, 
we now take the axioms of commutative quantum mechanics 
to apply with the simple replacement of $L^2$ by $\mathcal{H}_q$. 
Although this provides a consistent interpretational framework, 
the measurement of position needs more careful consideration. 
The problem with a measurement of position is not that 
the axioms above do not apply to the hermitian operators 
$\hat{X}$ and $\hat{Y}$, rather the problem is that these 
operators do not commute and thus a precise measurement 
of one of these observables leads to total uncertainty in the other. 
Yet, we would like to preserve the notion of position in the 
sense of a particle being localized around a certain point. 
The best that can be done in the noncommutative case is to 
construct a minimal uncertainty state in noncommutative 
configuration space and use that to give meaning to the 
notion of position. We now move on to describe this procedure.    

\noindent The minimal uncertainty states on noncommutative 
configuration space, which is isomorphic to boson Fock space, 
are well known to be the normalized coherent states \cite{klaud}
\begin{equation}
\label{cs} 
|z\rangle = e^{-z\bar{z}/2}e^{z b^{\dagger}} |0\rangle
\end{equation}
where, $z=\frac{1}{\sqrt{2\theta}}\left(x+iy\right)$ 
is a dimensionless complex number. These states provide an overcomplete 
basis on the noncommutative configuration space. 
Corresponding to these states we can construct a state 
(operator) in quantum Hilbert space as follows
\begin{equation}
|z, \bar{z} )=\frac{1}{\sqrt{\theta}}|z\rangle\langle z|.
\label{csqh}
\end{equation}
These states also have the property
\begin{equation}
B|z, \bar{z})=z|z, \bar{z}).
\label{p1}
\end{equation}
Writing the trace in terms of coherent states (\ref{cs}) and using 
$|\langle z|w\rangle|^2=e^{-|z-w|^2}$ it is easy to see that 
\begin{equation}
(z, \bar{z}|w, \bar{w})=\frac{1}{\theta}tr_{c}
(|z\rangle\langle z|w\rangle\langle w|)=
\frac{1}{\theta}|\langle z|w\rangle|^2=\frac{1}{\theta}e^{-|z-w|^2}
\label{p2}
\end{equation}
which shows that $|z, \bar{z})$ is indeed a Hilbert-Schmidt operator.  
We can now construct the `position' representation of a state 
$|\psi)=\psi(\hat{x},\hat{y})$ as
\begin{equation}
(z, \bar{z}|\psi)=\frac{1}{\sqrt\theta}tr_{c}
(|z\rangle\langle z| \psi(\hat{x},\hat{y}))=
\frac{1}{\sqrt\theta}\langle z|\psi(\hat{x},\hat{y})|z\rangle.
\label{posrep}
\end{equation}
In particular, introducing momentum eigenstates
\begin{eqnarray}
|p)=\sqrt{\frac{\theta}{2\pi}}e^{i\sqrt{\frac{\theta}{2}}
(\bar{p}b+pb^\dagger)}
\label{eg}
\end{eqnarray}
satisfying
\begin{eqnarray}
\hat{P}_{x}|p)=p_{x}|p)\quad,\quad\hat{P}_{y}|p)=p_{y}|p)
\label{eg1}
\end{eqnarray}
\begin{eqnarray}
(p'|p)=e^{-\frac{\theta}{4}(\bar{p}p+\bar{p}'p')}
e^{\frac{\theta}{2}\bar{p}p'}\delta(p-p')
\label{eg2}
\end{eqnarray}
we have
\begin{eqnarray}
(z, \bar{z}|p)=\frac{1}{\sqrt{2\pi}}
e^{-\frac{\theta}{4}\bar{p}p}e^{i\sqrt{\frac{\theta}{2}}(p\bar{z}+\bar{p}z)}~.
\label{eg3}
\end{eqnarray}
We now proceed to obtain the completeness relations for the momentum
and position eigenstates ($|p)$ and $|z,\bar{z})$) which are important
ingredients in the construction of the path
integral representation. To do this, using (\ref{eg3}), we compute
\begin{eqnarray}
\int d^{2}p(w, \bar{w}|p)(p|z, \bar{z})=\frac{1}{\theta}e^{-|w-z|^2}
=(w, \bar{w}|z, \bar{z})
\label{eg4}
\end{eqnarray}
which implies that the momentum eigenstates $|p)$ satisfy the following
completeness relation
\begin{eqnarray}
\int d^{2}p~|p)(p|=1_{Q}~.
\label{eg5}
\end{eqnarray}
The position eigenstates $|z,\bar{z})$, on the other hand, satisfy the
following completeness relation
\begin{eqnarray}
\int \frac{\theta dzd\bar{z}}{2\pi}~|z, \bar{z})\star(z, \bar{z}|=1_{Q}
\label{eg6}
\end{eqnarray}
where the star product between two functions 
$f(z, \bar{z})$ and $g(z, \bar{z})$ is defined as
\begin{eqnarray}
f(z, \bar{z})\star g(z, \bar{z})=f(z, \bar{z})
e^{\stackrel{\leftarrow}{\partial_{\bar{z}}}
\stackrel{\rightarrow}{\partial_z}} g(z, \bar{z})~.
\label{eg7}
\end{eqnarray}
To prove this, we use (\ref{eg3}) and compute
\begin{eqnarray}
\int \frac{\theta dzd\bar{z}}{2\pi}
(p'|z, \bar{z})\star(z, \bar{z}|p)=e^{-\frac{\theta}{4}(\bar{p}p+\bar{p}'p')}
e^{\frac{\theta}{2}\bar{p}p'}\delta(p-p')=(p'|p)~.
\label{eg8}
\end{eqnarray}
Thus, the position representation of the noncommutative
system maps quite naturally to the Moyal plane.
With the above formalism and the completeness relations for the
momentum and the position eigenstates 
(\ref{eg5}, \ref{eg6}) in place, we now proceed to write down the
path integral for the free particle propagation kernel
on the two dimensional noncommutative space. This reads
\begin{eqnarray}
(z_f, t_f|z_0, t_0)&=&\lim_{n\rightarrow\infty}\int \left(\frac{\theta}{2\pi}\right)^{n}
(\prod_{j=1}^{n}dz_{j}d\bar{z}_{j})~(z_f, t_f|z_n, t_n)\star_n
(z_n, t_n|....|z_1, t_1)\star_1(z_1, t_1|z_0, t_0)\nonumber\\
&=&\lim_{n\rightarrow\infty}\int_{-\infty}^{+\infty} \frac{1}{(2\pi)^n}
(\prod_{j=1}^{n}dx_{j}dy_{j})~(z_f, t_f|z_n, t_n)\star_n
(z_n, t_n|....|z_1, t_1)\star_1(z_1, t_1|z_0, t_0)~.
\label{pint1}
\end{eqnarray}
Now we compute the propagator over a small segment in the
above path integral. With the help of (\ref{eg3}) and (\ref{eg5}),
we have
\begin{eqnarray}
(z_{i+1}, t_{i+1}|z_i, t_i)&=&(z_{i+1}|e^{-iH\tau}|z_i)\nonumber\\
&=&(z_{i+1}|1-iH\tau +O(\tau^2)|z_i)\nonumber\\
&=&\int_{-\infty}^{+\infty}d^{2}p_i~(z_{i+1}|p_i)(p_i|z_i)
e^{-i\tau\frac{\vec{p}_{i}^2}{2m}}\nonumber\\
&=&N\exp{[-\beta(\vec{x}_{i+1}-\vec{x}_{i})^2]}
\label{pint2}
\end{eqnarray} 
where, $N=\frac{m}{m\theta+i\tau}$, $\beta=\frac{m}{2(m\theta+i\tau)}$
and $H=\frac{\vec{P_i}^2}{2m}$ being the Hamiltonian for the free
particle acting on the quantum Hilbert space. Using the above result,
we now write down after some algebra, the following generic 
element in the above path integral (\ref{pint1})
\begin{eqnarray}
\int_{-\infty}^{+\infty}dx_{i}dy_{i}~(z_{i+1},t_{i+1}|z_i,t_i)
\star_{i}(z_i,t_i|z_0,t_0)=N_{1}N_{2}
\left(\frac{\pi}{\beta_{1}\Lambda}\right)
\exp{\left[-\frac{\beta_{1}\gamma}{\Lambda}
(\vec{x}_{i+1}-\vec{x}_{0})^2\right]}
\label{pint3}
\end{eqnarray} 
where, $\gamma=\beta_{2}/\beta_{1}$ and $\Lambda=1+\gamma-2\theta\beta_{2}$.
It should be noted that the above computation have been carried out
with
\begin{eqnarray}
(z_{i+1},t_{i+1}|z_i,t_i)&=&N_{1}
\exp{[-\beta_1(\vec{x}_{i+1}-\vec{x}_{i})^2]}\nonumber\\
(z_{i},t_{i}|z_0,t_0)&=&N_{2}
\exp{[-\beta_2(\vec{x}_{i}-\vec{x}_{0})^2]}
\label{pint4}
\end{eqnarray} 
where, $N_1$, $N_2$ $\neq N$ and $\beta_1$, $\beta_2$ $\neq \beta$.
The reason for doing this will become clear as we proceed further.
We start by computing each of the integrals in (\ref{pint1}) 
from the extreme right. 

\noindent The first integral that we encounter on the extreme right
of the path integral (\ref{pint1}) is the following 
\begin{eqnarray}
\int_{-\infty}^{+\infty}dx_{1}dy_{1}~(z_{2},t_{2}|z_1,t_1)
\star_{1}(z_1,t_1|z_0,t_0)~.
\label{pint4a}
\end{eqnarray} 
To compute this integral, we set $i=1$ on the 
left hand side of (\ref{pint3}) 
and set $N_1=N_2=N$ and $\beta_1=\beta_2=\beta$ 
on the right hand side of (\ref{pint3}) which yields
\begin{eqnarray}
\int_{-\infty}^{+\infty}dx_{1}dy_{1}~(z_{2},t_{2}|z_1,t_1)
\star_{1}(z_1,t_1|z_0,t_0)=\frac{2\pi m}{(m\theta+2i\tau)}
\exp{\left[-\frac{\beta}{2(1-\theta\beta)}
(\vec{x}_{2}-\vec{x}_{0})^2\right]}~.
\label{pint5}
\end{eqnarray} 
The first two integrals starting from the extreme right of
(\ref{pint1}) reads the following
\begin{eqnarray}
\int_{-\infty}^{+\infty}(dx_{2}dy_{2})(dx_{1}dy_{1})~(z_{3},t_{3}|z_2,t_2)
\star_{2}(z_2,t_2|z_1,t_1)\star_{1}(z_1,t_1|z_0, t_0)~.
\label{pint6}
\end{eqnarray} 
Using (\ref{pint5}), we observe that the above integral (\ref{pint6})
is of the form (\ref{pint3}) with $i=2$ and $N_1=N$, 
$N_2=\frac{2\pi m}{(m\theta+2i\tau)}$, $\beta_1=\beta$ 
and $\beta_2=\frac{\beta}{2(1-\theta\beta)}$. Hence, we obtain
\begin{eqnarray}
\int_{-\infty}^{+\infty}(dx_{2}dy_{2})(dx_{1}dy_{1})~(z_{3},t_{3}|z_2,t_2)
\star_{2}(z_2,t_2|z_1,t_1)\star_{1}(z_1,t_1|z_0, t_0)
=\frac{(2\pi)^{2}m}{(m\theta+3i\tau)}
\exp{\left[-\frac{\beta}{(3-4\theta\beta)}
(\vec{x}_{2}-\vec{x}_{0})^2\right]}~.
\label{pint6a}
\end{eqnarray} 
Repeating this procedure $n$ times, we finally obtain the free particle
propagation kernel on the two dimensional noncommutative space
\begin{eqnarray}
(z_f, t_f|z_0, t_0)&=&\lim_{n\rightarrow\infty}
\frac{m}{[m\theta+i(n+1)\tau]}
\exp{\left[-\frac{\beta}{[(n+1)-2n\theta\beta]}
(\vec{x}_{f}-\vec{x}_{0})^2\right]}\nonumber\\
&=&\lim_{n\rightarrow\infty}\frac{m}{[m\theta+i(n+1)\tau]}
\exp{\left[-\frac{m}{2[i(n+1)\tau+m\theta]}
(\vec{x}_{f}-\vec{x}_{0})^2\right]}\nonumber\\
&=&\frac{m}{(m\theta+iT)}
\exp{\left[-\frac{m}{2(iT+m\theta)}
(\vec{x}_{f}-\vec{x}_{0})^2\right]}\quad;\quad (n+1)\tau=T=t_{f}-t_{0}~.
\label{pint7}
\end{eqnarray}

\noindent {\it{Conclusion}} :\\
 
\noindent In this paper, we have formulated the path integral
representation of the free particle propagation kernel based on 
the newly established consistent formulation of noncommutative
quantum mechanics. In contrast to the approach in 
\cite{spallucci}, where the star product does not arise, 
we have shown that the star product plays an important role in 
our approach and interestingly, the star product still
allows the exact computation of the free particle propagator
to all orders in the noncommutative parameter $\theta$. 
The result for the propagator exhibits the ultra-violet
cutoff induced by the noncommuttaive parameter and is in conformity
with the result obtained earlier in the literature \cite{spallucci}.\\

\noindent {\bf{Acknowledgements}} : This work was supported under a grant of the National Research Foundation of South Africa. 


\end{document}